\documentclass{aaStyle/aa}

\usepackage{graphicx}
\usepackage{natbib}
\usepackage{amsmath}
\usepackage[english]{babel}

\newcommand{\Abst}[1]{\,#1}

\newcommand{\kB}{k_{\rm B}}
\newcommand{\me}{m_{\rm e}}
\newcommand{\Ne}{n_{\rm e}}
\newcommand{\Te}{T_{\rm e}}
\newcommand{\sigT}{\sigma_{\rm T}}
\newcommand{\rC}{r_{\rm c}}

\newcommand{\beqa}{\begin{eqnarray}}   
\newcommand{\eeqa}{\end{eqnarray}}   
\newcommand{\bsub}{\begin{subequations}}
\newcommand{\esub}{\end{subequations}}

\newcommand{\AKc}[1]{{\rm C}_{#1}}
\newcommand{\AKs}[1]{{\rm S}_{#1}}
\newcommand{\vek}[1]{\mbox{\boldmath${#1}$\unboldmath}}
\begin{document}

\title{Kinetic Sunyaev-Zeldovich effect from galaxy cluster rotation}
\titlerunning{Kinetic SZ effect from galaxy cluster rotation}

\author{J. Chluba\inst{1,3} 
\and K. Mannheim\inst{2,3}}
\authorrunning{J. Chluba and K. Mannheim}

\offprints{\\J. Chluba, \email{jchluba@mpa-garching.mpg.de}}

\institute{Max-Planck-Institut f\"ur Astrophysik, Karl-Schwarzschild-Str. 1, 85740
Garching bei M\"unchen, Germany
\\e-mail: jchluba@mpa-garching.mpg.de
%-----------------
\and
%-----------------
Institut f\"ur Theoretische Physik und Astrophysik, Am Hubland, 97074
W\"urzburg, Germany
\\ e-mail: mannheim@astro.uni-wuerzburg.de
%-----------------
\and
%-----------------
Universit\"ats-Sternwarte G\"ottingen, Geismarlandstr. 11, 37083
G\"ottingen, Germany}
%-----------------

\date{Received; Accepted }

\abstract {We show how the temperature and the polarisation of the cosmic
microwave background are affected by bulk rotation of clusters of galaxies
owing to the kinetic Sunyaev-Zeldovich effect.  The main effects of rotation
are (i) a shift of the position of the peak of the temperature fluctuation
relative to the center of the cluster by a few percent of the core radius and
(ii) a tilt of the direction of the plane of linear polarisation by several
degrees.
%-----------------
}

\maketitle

\section{Introduction} 
Several effects lead to anisotropies of the cosmic microwave background (CMB):
primary effects, imprinted on the surface of last scattering, and secondary
effects, arising after hydrogen-recombination or after reionisation has taken
place \citep[for references see][]{White2002}.

Among the secondary effects, the thermal Sunyaev-Zeldovich effect (th-SZE),
which is due to inverse Compton scattering of the CMB photons off the hot
intracluster medium (ICM) \citep{SunyZeld69}, and the kinetic
Sunyaev-Zeldovich effect (k-SZE), which arises from the peculiar motion of the
cluster in the rest frame of the CMB \citep{SunyZeld80}, are most
important. The cosmological importance of the th-SZE is due to the fact that
it is redshift independent. Therefore it can be used to detect clusters of
galaxies at redshifts, where other observational methods fail
\citep{Korolev86,Kneissl2001}, and to determine the Hubble constant
\citep{Birkinshaw1999, Reese2002}.
%-------------
In principle the k-SZE can be used to determine both components of the
peculiar velocity: the line of sight component via the relative change of
intensity of the CMB and the velocity component in the celestial plane via the
degree of linear polarisation of the CMB radiation. Measurements of the k-SZE
are extremely difficult and will probably become feasible in the near future.

In this paper we address the k-SZE arising from bulk rotation of the ICM. In
order to distinguish this effect from the k-SZE due to a bulk translatory
motion of the cluster with respect to the CMB, we shall henceforth coin it the
rotational kinetic Sunyaev-Zeldovich effect (rk-SZE). To our knowledge, the
rk-SZE was previously only considered by \cite{Cooray2002}. In their work the
rk-SZE was discussed for a gas density profile following from hydrostatic
equilibrium of the gas in the Navarro-Frenk-White dark matter density field
within a halo \citep{NFW96, Makino98}. Assuming isothermality this gas density
profile is very well approximated by the commonly used isothermal
$\beta$-model \citep{Caval76}, which is better applicable to analytical
calculations. Here we focus on the contributions of the rk-SZE to the
temperature fluctuations and the linear polarisation of the CMB for the
isothermal $\beta$-model finding a new method to measure the rotational
properties of a cluster by performing multifrequency measurements of the
position of the peak of the temperature fluctuation.

The paper is organized as follows. In Sect. \ref{sec:calc} we state the model
assumptions for the rotating cluster of galaxies and derive analytic formulae
describing the rk-SZE. We also derive formulae describing combined thermal,
kinetic and rotational kinetic SZE. In Sect. \ref{sec:dis} we discuss our
analytic results using recent observational data to estimate the effects for a
set of 18 clusters in the redshift range from $z\sim 0.14$ to $z\sim
0.78$. Finally, we draw our conclusions in Sect. \ref{sec:conc}.

\newpage
\section{Calculations} 
\label{sec:calc}
%----------------
\subsection{General equations}
\subsubsection*{Kinetic Sunyaev-Zeldovich effect}
The relative change of intensity of the CMB due to the k-SZE of a small volume
of electron gas with a line of sight optical depth $d\tau=\sigT\,\Ne\,dl$,
where $\sigT$ is the Thomson cross section and $\Ne$ is the electron density,
is given as \citep{SunyZeld80}
%-----------------------------
\beqa\label{eq:DIntkinSZE}
d\left(\frac{\Delta I_{\nu}}{I_{\nu}}\right)=
G_{\nu}\,\beta_{\parallel}\,d\tau
\Abst{.}
\eeqa
%-----------------------------
Here $\beta_{\parallel}$ is the line of sight velocity in units of the speed
of light $c$ and $G_\nu=x\,e^x/(e^x - 1)$ with the dimensionless frequency
$x=h\nu/\kB T$, where $T$ is the temperature of the photon field, $h$ is the
Planck constant and $\kB$ is the Boltzmann constant. $\beta_{\parallel}$ is
defined to be positive if the cluster is approaching the observer.
%-----------------------------
The degree of linear polarisation of the CMB due to the k-SZE can be
written as \citep{SunyZeld80}
%---------------
\beqa\label{eq:DPolRJ}
dP=\frac{1}{10}\,G_{\nu,\rm pol}\,\beta^2_{\perp}\,d\tau
\Abst{,}
\eeqa
%---------------
where now $\beta_{\perp}$ is the velocity component perpendicular to the line
of sight and $G_{\nu,\rm pol}$ is given as
%---------------
\beqa\label{eq:GPOLX}
G_{\nu,\rm pol}=\frac{e^x\,(e^x+1)}{2\,(e^x-1)^2}\,x^2
\Abst{.}
\eeqa
%---------------
The plane of polarisation lies perpendicular to the direction of
$\beta_{\perp}$. In the Rayleigh-Jeans part of the CMB spectrum $G_{\nu,\rm
pol}$ approaches unity and it increases as $\sim x^2/2$ in the Wien tail
\citep{SunyZeld80, Saz99, Aud99}.

\subsubsection*{Thermal Sunyaev-Zeldovich effect}
The relative change of intensity of the CMB due to the th-SZE of a
cluster of galaxies is given as
\citep{SunyZeld69}
%--------------------
\beqa\label{eq:thSZIntens}
\frac{\Delta I_{\nu}}{I_{\nu}} = F_\nu\cdot y_{\rm C}
\Abst{,}
\eeqa
%--------------------
where $F_\nu$ and the Compton parameter $y_{\rm C}$ are defined as
%--------------------
\beqa
\label{eq:FX}
F_\nu&=&\frac{x e^x}{e^x-1}\left[x\frac{e^x+1}{e^x-1} -4 \right] ,\\
\label{eq:ComptonPar}
y_{\rm C} &=&\int \theta_{\rm e}\,\sigT n_{\rm e}\,dl 
\eeqa 
%--------------------
with the dimensionless temperature of the electron gas, $\theta_{\rm
e}=\kB\Te/\me c^2$, where $\me$ is the rest mass of an electron. The typical
temperature $\Te$ of the electron gas inside clusters is $\sim
5\,\text{keV}$. Relativistic corrections thus lead to small changes in the
form of $F_\nu$ which are neglected in the following \citep[for details
see][]{Itoh98, Chall98,Ensslin2000}. In the Rayleigh-Jeans part of the CMB
spectrum $F_\nu$ approaches $-2$ and it increases as $\sim x^2$ in the Wien
tail. At the crossover frequency $x_{\rm cr}\sim 3.83$ the th-SZE vanishes and
the k-SZE reaches its maximum.

\subsection{Model of a rotating cluster of galaxies}
We describe the electron density profile of a rotating cluster of galaxies
with an isothermal $\beta$-model \citep{Caval76}. In order to account for
asphericities, we introduce the parameter $\iota=\rC/r_{\rm z'}$, where $\rC$
is the core radius of the cluster and $r_{\rm z'}$ is the distance in the
direction $z'$, at which the electron density is equal to the density at the
core radius in the spherical case. For $\iota < 1$ the cluster is prolate, for
$\iota > 1$ it is oblate and spherical in the case $\iota=1$. The modified
isothermal $\beta$-model then, in the main axis system $S'$, has the form
%---------------
\beqa\label{eq:rotClustDichInk}
\Ne(\vek{r})=n_{\rm e0}\left(1+
\frac{{x'}^2+{y'}^2+\iota^2 {z'}^2}{r^2_{\rm c}}\right)^{-\gamma}
\Abst{,}
\eeqa
%---------------
where $n_{\rm e0}$ is the central electron density and $\gamma=3\beta/2$, with
a typical value $\beta=2/3$.

Very little, if anything, is known about the rotational properties of
large-scale structures such as clusters of galaxies. But the existence of an
angular momentum distribution in large scale structures is very likely:
Self-gravitating flows generally produce rotating structures on small scales
(e.g.  spiral galaxies, stars), even if the progenitor structure has a
vanishing net angular momentum. Flattened geometries of clusters of galaxies
may be a hint, that an angular momentum profile exists on larger scales as
well. Recent numerical simulation of off-center cluster mergers predict, that
a significant angular momentum associated with velocities up to few
$10^{3}\,\text{km/s}$ can be induced \citep{Roettiger2000,
Roettiger1998}. Also first observational evidence for an angular momentum
distribution in clusters was found by measuring the Doppler shift of X-ray
spectral lines \citep{Dupke2001}. In recent simulations \cite{Bullock2001}
found first evidence for an universal angular momentum profile in dark matter
halos, which is consistent with solid body rotation with typical rotational
velocity of $5\,\%$ of the circular velocity. Assuming that the gas follows
the dark matter motion, we therefore use solid body rotation to describe the
velocity field of the ICM \citep[Further reasons are given in][]{Cooray2002}.

%---------------
%Since there is currently no detailed information related to cluster gas
%rotation available -- neither observationally nor numerically -- we assume
%solid body rotation as a simple model for the motion of the electron gas in
%the cluster. In a paper of \citealt{Bullock2001}, first evidence for a
%universal angular momentum profile for galactic halos was found in numerical
%simulations which is consistent with solid body rotation and a typical
%rotational velocity of $5\,\%$ of the circular velocity. In
%\citealt{Cooray2002} it was also argued that this is a rather good
%assumption. At least for estimates of the effects it is a useful starting
%point.
%---------------

\subsection{Kinetic Sunyaev-Zeldovich effect from cluster rotation}
Equation \eqref{eq:DIntkinSZE} can be used to calculate the relative intensity
change of the CMB due to the internal motion of the cluster medium. If we
assume that the angular velocity vector \vek{\omega} is anti-parallel to the
$z'$-axis and we define the frame $S$ of an observer such that the $z$-axis
lies along the line of sight and the $x$-axis lies along the projection of
\vek{\omega} onto the celestial plane, then the line of sight velocity is
given as $\beta_{\parallel}=\beta_{\rm c}\,y\,\AKs{i}/ \rC$, where $\beta_{\rm
c}=\omega\,\rC/c$ is the rotational velocity at the core radius and $\AKs{i}$
denotes the sine of the inclination $i$ of the observer to the rotational
axis.
%---------------
The derivation of the relative change of intensity for our simple model of a
rotating cluster then reduces to the calculation of the optical depth as a
function of $x$ and $y$, which is given in the appendix
\ref{sec:opticaldepth}. It follows 
%-------------
\beqa\label{eq:ResultkinRotClusterInk} \left(\frac{\Delta
I_{\nu}}{I_{\nu}}\right)_{\rm rot} 
=G_{\nu}\,\beta_{\rm c} 
\,\frac{y\,\AKs{i}}{\rC}\cdot \tau(x,y,i,\gamma)
\Abst{,}
\eeqa
%-------------
where the definition of the optical depth $\tau(x,y,i,\gamma)$ is given in the
appendix (Eq. \ref{eq:TAU}). From Eq. \eqref{eq:ResultkinRotClusterInk} it is
clear, that there is no effect for an inclination $i=0$ and it is maximal for
$i=\pi/2$.

\begin{figure}
\resizebox{\hsize}{!}{\includegraphics{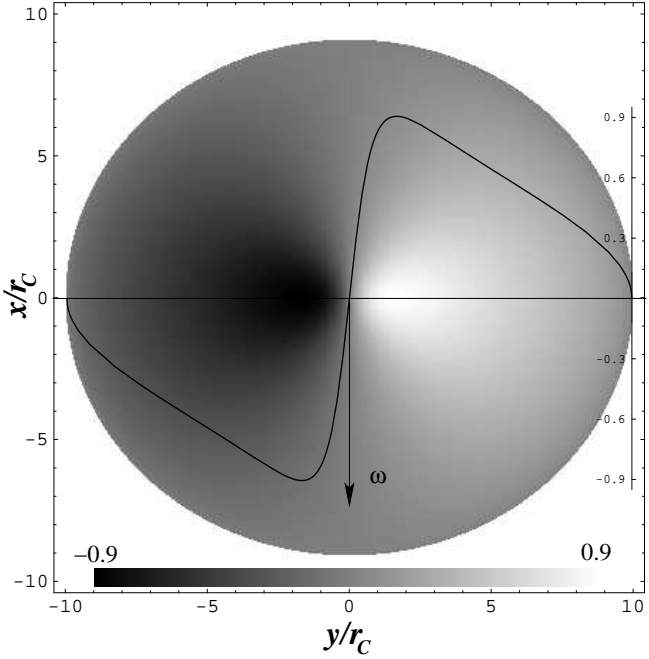}}
\caption{Relative change of intensity (grayplot) in units of $G_{\nu}\,\beta_{\rm
c}\,\tau_{\rm c}$ due to the rk-SZE for an oblate cluster ($\iota=1.1$) with a
cutoff radius $R=10\,\rC$, an inclination $i=\pi/2$ and $\gamma=1.125$. The
overlayed plot (solid line) shows a cut along $x=0$.}
\label{fig:kinRotClusterInk}
\end{figure}

In Fig. \ref{fig:kinRotClusterInk} the relative change of intensity due to the
rk-SZE is shown for an oblate cluster $(\iota=1.1)$ with a cutoff radius
$R=10\,\rC$, an inclination $i=\pi/2$ and $\gamma=1.125$. The distinct
dipolar-like pattern arises from the fact that one part of the gas is moving
towards and the other part is moving away from the observer.  As can be seen
in the overlayed plot, the peak value of the relative intensity change is
about $\sim 0.9\,\tau_{\rm c}\,\beta_{\rm c}\,G_{\nu}$. For further discussion
see Sect. \ref{sec:dis}.
%----------------
%----------------

\subsection{Superposition of all the Sunyaev-Zeldovich contributions}
%---------------------
In general the relative change of intensity of the CMB has contributions from
the thermal, the kinetic and the rotational kinetic SZE. Assuming that the
cluster of galaxies is moving with a line of sight velocity
$\beta_{\parallel,\rm b}$ one finds with Eqns. \eqref{eq:DIntkinSZE},
\eqref{eq:thSZIntens} and \eqref{eq:ResultkinRotClusterInk}
%-------------
\beqa\label{eq:RELCHANGETOT} 
\frac{\Delta I_{\nu}}{I_{\nu}} 
=
\left(F_{\nu}\,\theta_{\rm e} + 
G_{\nu}\left[
\beta_{\parallel,\rm b}+\beta_{\rm c}\,\frac{y\,\AKs{i}}{\rC}
\right]
\right)
\cdot \tau(x,y,i,\gamma)
\eeqa
%-------------
for our simple model of a rotating cluster as described above. Depending on
the observed frequency $\nu$ the importance of the different contributions
varies. The th-SZE dominates in the Rayleigh-Jeans and Wien part of the CMB
spectrum, whereas the k-SZE and the rk-SZE dominate around the crossover
frequency $\nu_{\rm c} \sim 218\,\text{GHz}$. If the cluster is not rotating,
the absolute value of relative change of intensity reaches its maximum at the
center of the cluster, where the line of sight optical depth is highest. If
there is only a contribution of the rk-SZE the relative change of intensity
has two distinct extrema symmetrically around the center of the cluster.

It is possible to find the positions of the extrema of the relative change of
intensity analytically from Eq. \eqref{eq:RELCHANGETOT}, if one assumes
$R\rightarrow\infty$ and $\gamma>\frac{1}{2}$. For $\gamma\le\frac{1}{2}$ the
density profile decreases slower than $\sim 1/r$ and therefore leads to
infinite optical depth for $r\rightarrow\infty$. Although $R\rightarrow\infty$
implies superluminal rotation speeds far away from the center of the cluster,
these unphysical contributions to the rk-SZE are small for typical values of
$\gamma$ due to the rapidly, radially decreasing density (see
fig. \ref{fig:kinRotClusterInkmax} and the discussion).

%-------------
Defining the parameter $\eta$ as
%-------------
\beqa\label{eq:ETA} 
\eta=
\frac{2\,G_{\nu}\,\beta_{\rm c}}{F_{\nu}\,\theta_{\rm e} 
+\beta_{\parallel,\rm b}\,G_{\nu}}
\eeqa
%-------------
and using Eqns. \eqref{eq:RELCHANGETOT} and \eqref{eq:IntIinfty} one finds the
formal solution
%----------
\beqa\label{eq:Versch}
y_{\rm extr}=
-\frac{\rC\,(2\gamma-1)}{2\eta\AKs{i}(\gamma-1)}
\left[1\pm \sqrt{1+\frac{2\eta^2\AKs{i}^2(\gamma-1)}{(2\gamma-1)^2}}\,\,\right]
\eeqa
%----------
for the positions of the extrema relative to the center. For
$\gamma\rightarrow 1$ and $\eta\ne 2\,G_{\nu}\,\beta_{\rm c}$ there is only
one extremum at
%----------
$|y_{\rm extr}|=|\eta\AKs{i}|\,\rC/2$.
%----------
In the case $\frac{1}{2}<\gamma< 1$ the value of $\eta\AKs{i}$ is limited to
the region $|\eta\AKs{i}|\leq \zeta$, where $\zeta$ is defined as
%----------
\beqa\label{eq:zeta}
\zeta=\frac{2\gamma-1}{\sqrt{2|\gamma-1|}}
\Abst{.}
\eeqa
%----------
For $\gamma\ge 1$ the value of $\eta\AKs{i}$ is arbitrary. 

Equation \eqref{eq:Versch} has the limiting forms ($\gamma\ne 1$)
%----------
\beqa\label{eq:ApprVersch1}
y_{\rm extr}\approx
\mp\frac{\eta\AKs{i}\,\rC}{2(2\gamma-1)}\cdot
\left[1+\frac{(1\pm 1) (2\gamma-1)^2}{\eta^2\AKs{i}^2 (\gamma-1)}\right]
\eeqa
%----------
for $\gamma>\frac{1}{2}$ and $|\eta\AKs{i}|\ll \zeta$ and
%----------
\beqa\label{eq:ApprVersch2}
y_{\rm extr}\approx
\pm\frac{\rC}{\sqrt{2(\gamma-1)}}
\eeqa
%----------
for $\gamma>1$ and $|\eta\AKs{i}|\gg \zeta$. There are no extrema for
$\gamma\le 1$ and $|\eta\AKs{i}|\gg \zeta$.
 
The first case, i.e. Eq. \eqref{eq:ApprVersch1}, corresponds to a small
contribution of the rk-SZE to the relative change of intensity compared to the
thermal and the kinetic SZE. In this case there is an extremum at
%----------
$y_{\rm extr}\sim\eta\AKs{i}\,\rC/2(2\gamma-1)$ 
%----------
near the center of the cluster and one far away from the center. The second
case, i.e. Eq. \eqref{eq:ApprVersch2}, corresponds to a rk-SZE with negligible
contributions of the thermal and the kinetic SZE. See Sect. \ref{sec:dis} for
further discussion.

\subsection{Polarisation due to the rotational kinetic Sunyaev-Zeldovich
effect} 
%---------------------
Polarized radiation can be described by the Stokes parameters
$I,\,Q,\,U,\,V$. The $Q$ and $U$ parameters describe the degree of linear
polarisation the $V$ parameter describes the degree of circular polarisation
\cite[for further details see][]{Rybicki79}. We assume that the incident CMB
radiation is isotropic and unpolarized ($Q=U=V=0$) and that the scattering
event can be treated as Thomson scattering (this means that the peculiar
velocity $\beta_{\rm p}$ of the scattering electron obeys $\beta_{\rm p} \ll
1$ and $h\nu \ll \me c^2$. The latter is valid for CMB photons). Since Thomson
scattering does not produce circular polarisation we will drop the
$V$-parameter in the following.

To calculate the contribution of the rk-SZE to the degree of linear
polarisation we have to find the scattered Stokes parameters $I^{\rm
s},\,Q^{\rm s},\,U^{\rm s}$ in the rest frame of the observer. The degree of
linear polarisation is then given by
%---------------
\beqa\label{eq:defPol}
P=\frac{\sqrt{(Q^{\rm s})^2+(U^{\rm s})^2}}{I^{\rm tot}}
\Abst{,}
\eeqa
%---------------
where the total intensity $I^{\rm tot}$ is given as the sum of the incident
intensity $I_0$, the scattered and the absorbed intensity: $I^{\rm
tot}=I_0\,(1-\tau)+I^{\rm s}$. The angle $\phi$ between the $x$-axis and the
plane of polarisation is given by
%---------------
\beqa\label{eq:defangle} 
\tan 2\phi=\frac{U^{\rm s}}{Q^{\rm s}}
\Abst{.}  
\eeqa
%---------------
%For $U^{\rm s}=0$ the plane of polarisation lies in $x$-direction. 
Since the scattered Stokes parameters $Q^{\rm s}$ and $U^{\rm s}$ are
proportional to $\tau$, consistent up to first order in $\tau$, the inverse of
the total intensity in Eq. \eqref{eq:defPol} can be approximated as $1/I^{\rm
tot}\approx 1/I_0$.

Assuming that the incident intensity $I_0$ is isotropic, the scattered Stokes
parameters $dQ^{\rm s}$ and $dU^{\rm s}$ in the observer frame for a small
volume of electron gas with optical depth $d\tau$ and peculiar velocity
$\beta_{\rm p}$ up to second order in $\beta_{\rm p}$ are given as
\citep{Saz99}
%---------------
\bsub\label{eq:scatteredStokes}
\beqa
\label{eq:scatteredQ}
dQ^{\rm s}&=&
\,\,\,\,
\frac{1}{10}\,I_0\,G_{\nu,\rm pol}\,\beta_{\perp,\rm p}^2\,\cos(2\chi)\,d\tau
\\[1mm]
\label{eq:scatteredU}
dU^{\rm s}&=&
-\frac{1}{10}\,I_0\,G_{\nu,\rm pol}\,\beta_{\perp,\rm p}^2\,\sin(2\chi)\,d\tau
\Abst{,}
\eeqa
\esub
%---------------
where $\beta_{\perp,\rm p}$ is the velocity component in the celestial plane
and $\chi$ is the angle between $\beta_{\perp,\rm p}$ and the $x$-axis of the
observer.

For the model of a rotating cluster as described above the following relations
can be found
%---------------------
\bsub\label{eq:cossinrot}
\beqa 
\beta_{\perp,\rm r}^2 \cos(2\chi_{\rm r})
&=&
\,\,\,\,\,
\beta_{\rm c}^2
\left[
\tilde{y}^2 \AKc{i}^2 - \left(\tilde{x}\,\AKc{i}+\tilde{z}\,\AKs{i}\right)^2
\right]
\\[1mm]
\beta_{\perp,\rm r}^2 \sin(2\chi_{\rm r})
&=&-\beta_{\rm c}^2\,
\tilde{y}\,\AKc{i}
\left(
\tilde{x}\,\AKc{i} + \tilde{z}\,\AKs{i}
\right)
\Abst{,}
\eeqa
\esub
%---------------------
where $\beta_{\perp,\rm r}$ is the rotational velocity component in the
celestial plane, $\chi_{\rm r}$ is the angle between $\beta_{\perp,\rm r}$ and
the $x$-axis and $\tilde{x}$ and $\tilde{y}$ are the $x,y$-coordinates in
units of the core radius $\rC$. Inserting this into Eqns.
\eqref{eq:scatteredQ} and \eqref{eq:scatteredU} and integrating along the line
of sight we find
%---------------------
\bsub\label{eq:resultQUrot}
\beqa 
Q^{\rm s}_{\rm r}&=&
\kappa\,\beta_{\rm c}^2
\Big[
\left(\tilde{y}^2-\tilde{x}^2\right)\AKc{i}^2 \cdot\tau
- \AKs{i}^2\cdot J
-2\,\tilde{x}\,\AKc{i}\AKs{i}\cdot H
\Big]\\[1mm]
U^{\rm s}_{\rm r}&=&
2\,\kappa\,\beta_{\rm c}^2\,
\tilde{y}\,\AKc{i}\left[ \tilde{x}\,\AKc{i}\,
\cdot\tau
+ \AKs{i}\cdot H 
\right]
\Abst{,}
\eeqa
\esub
%---------------------
where $\kappa=\frac{1}{10}\,I_0\,G_{\nu,\rm pol}$ and the optical depth
$\tau(x,y,i,\gamma)$ and integrals $J(x,y,i,\gamma)$ and $H(x,y,i,\gamma)$ are
defined in the appendix (Eq. \ref{eq:TAU}, \ref{eq:DefJ} and
\ref{eq:DefH}). Now, with Eq. \eqref{eq:defPol} the degree of linear
polarisation due to the rk-SZE up to second order in $\beta_{\perp,\rm r}$
follows as
%---------------------
\beqa \label{eq:resultPOLrot}
P_{\rm r}&=&
\frac{1}{10}\,G_{\nu,\rm pol}\,\beta_{\rm c}^2
\Bigg[
\Big(\tilde{r}^2\AKc{i}^2 \cdot\tau
+\AKs{i}\left[\AKs{i}\cdot J+2\tilde{x}\,\AKc{i}\cdot H \right]
\Big)^2
\Bigg.
\nonumber\\[1mm]
& & \quad 
\Bigg. 
+4\,\tilde{y}^2\AKs{i}^2\AKc{i}^2 \cdot\Big(H^2-\tau\cdot J\Big)
\Bigg]^{\frac{1}{2}}
\Abst{.}
\eeqa
%---------------------
Here $\tilde{r}$ is the distance from the center of the cluster in units of
the core radius $\rC$. The direction of the planes of polarisation can be
found with Eq. \eqref{eq:defangle}.

%\subsubsection*{Case $i=0$}
%---------------------
For an inclination $i=0$ with Eq. \eqref{eq:resultPOLrot} one finds
%-------------
\beqa\label{eq:Poli0} 
P_{i=0,\rm r} 
=\frac{1}{10}\,G_{\nu,\rm pol}\,\beta^2_{\rm c}\,
\tilde{r}^2\cdot \tau(x,y,0,\gamma)
\eeqa
%-------------
for the degree of linear polarisation. The planes of polarisation lie along
the radial vector. This can be also found by inserting $\beta_{\perp,i=0,\rm
r}=\beta_{\rm c}\,\tilde{r}$ into Eq. \eqref{eq:DPolRJ} and integrating along
the line of sight.

\begin{figure}
\resizebox{\hsize}{!}{\includegraphics{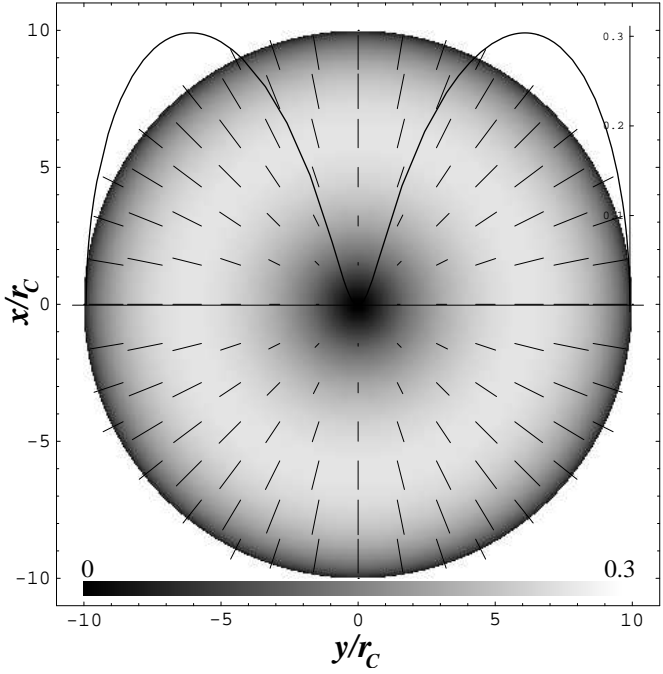}}
\caption{Degree of linear polarisation (grayplot) in units of $G_{\nu}\,\beta^2_{\rm
c}\,\tau_{\rm c}$ for a rotating, oblate cluster ($\iota=1.1$) with cutoff
radius $R=10\,\rC$, an inclination $i=0$ and $\gamma=1.125$. The short lines
indicate the direction of the planes of polarisation and their length the degree of polarisation. The overlayed plot (solid line) shows a cut along $x=0$.}
\label{fig:Poli0}
\end{figure}

The degree of linear polarisation due to the rk-SZE is shown in Fig.
\ref{fig:Poli0} for an oblate cluster $(\iota=1.1)$ with a cutoff radius
$R=10\,\rC$ and $\gamma=1.125$. The short lines indicate the direction of the
planes of polarisation and their length the degree of polarisation. As can be
seen in the overlayed plot, the degree of polarisation falls off steeply
towards the edge of the cluster.
%----------------
%----------------

%\subsubsection*{Case $i=\pi/2$}
%---------------------
For an inclination $i=\pi/2$ again with Eq. \eqref{eq:resultPOLrot} one finds
%-------------
\beqa\label{eq:Poli90} 
P_{i=\pi/2,\rm r} 
=\frac{1}{10}\,G_{\nu,\rm pol}\,\beta^2_{\rm c}
\cdot J(x,y,\pi/2,\gamma)
\eeqa
%-------------
for the degree of linear polarisation. The planes of polarisation lie along
the $x$-axis. Alternatively this result can be found by inserting
$\beta_{\perp,i=\pi/2,\rm r}=\beta_{\rm c}\,\tilde{z}$ into Eq.
\eqref{eq:DPolRJ} and integrating along the line of sight.

\subsection{Polarisation due to combined kinetic and rotational kinetic 
Sunyaev-Zeldovich effect}
%---------------------
If the cluster is rotating and in addition moving with the bulk translatory
velocity $\beta_{\perp,\rm b}$ in the celestial plane one can find the relations
%---------------------
\bsub\label{eq:cossinsup}
\beqa 
\beta_{\perp}^2 \cos(2\chi)
&=&\beta_{\perp,\rm b}^2 \cos(2\alpha) 
+\beta_{\perp,\rm r}^2 \cos(2\chi_{\rm r})
\nonumber\\[1mm]
& &  
\!\!\!
-2\beta_{\perp,\rm b}\,\beta_{\rm c}
\Big[
\left(
\tilde{y}\,\AKc{\alpha}
+\tilde{x}\,\AKs{\alpha}
\right)\AKc{i}
+\tilde{z}\,\AKs{\alpha}\AKs{i}
\Big]
\\[1mm]
\beta_{\perp}^2 \sin(2\chi)
&=&\beta_{\perp,\rm b}^2 \sin(2\alpha)
+\beta_{\perp,\rm r}^2 \sin(2\chi_{\rm r})
\nonumber\\[1mm]
& &  
\!\!\!
-2\beta_{\perp,\rm b}\,\beta_{\rm c}
\Big[
\left(
\tilde{y}\,\AKs{\alpha}
-\tilde{x}\,\AKc{\alpha}
\right)\AKc{i}
-\tilde{z}\,\AKc{\alpha}\AKs{i}
\Big]
,
\eeqa
\esub
%---------------------
where $\beta_{\perp}$ is the total velocity in the celestial plane, $\chi$ and
$\alpha$ are the angles between the $x$-axis and $\beta_{\perp}$ and
$\beta_{\perp, \rm b}$ respectively and the definitions \eqref{eq:cossinrot}
were used. Now with Eqns. \eqref{eq:scatteredQ} and \eqref{eq:scatteredU}
we find
%---------------------
\bsub\label{eq:resultQUsup}
\beqa 
Q^{\rm s}&=&
Q^{\rm s}_{\rm r}
+\kappa\,\beta_{\perp,\rm b}^2 \cos(2\alpha)\cdot\tau
\nonumber\\[1mm]
& & \!\!
-2\,\kappa\,\beta_{\perp,\rm b}\,\beta_{\rm c}
\Big[
\left(
\tilde{y}\,\AKc{\alpha}
+\tilde{x}\,\AKs{\alpha}
\right)\AKc{i}\cdot\tau
+\AKs{\alpha}\AKs{i}\cdot H
\Big]
\\[1mm]
U^{\rm s}&=&
U^{\rm s}_{\rm r}
-\kappa\,\beta_{\perp,\rm b}^2 \sin(2\alpha)\cdot\tau
\nonumber\\[1mm]
& & \!\! 
+2\,\kappa\,\beta_{\perp,\rm b}\,\beta_{\rm c}
\Big[
\left(
\tilde{y}\,\AKs{\alpha}
-\tilde{x}\,\AKc{\alpha}
\right)\AKc{i}\cdot\tau
-\AKc{\alpha}\AKs{i}\cdot H
\Big]
.
\eeqa
\esub
%---------------------
%The degree of linear polarisation and the directions of the planes of
%polarisation can be found with Eqns. \eqref{eq:defPol} and
%\eqref{eq:defangle}. 
%---------------------
To demonstrate the effects of the rk-SZE on the degree of
linear polarisation of the CMB, we now consider the two special cases $i=0$
and $i=\pi/2$ with an additional bulk translatory motion of the cluster along
the $y$-axis.

\subsubsection*{Case $i=0$ and $\alpha=\pi/2$}
\begin{figure}
\resizebox{\hsize}{!}{\includegraphics{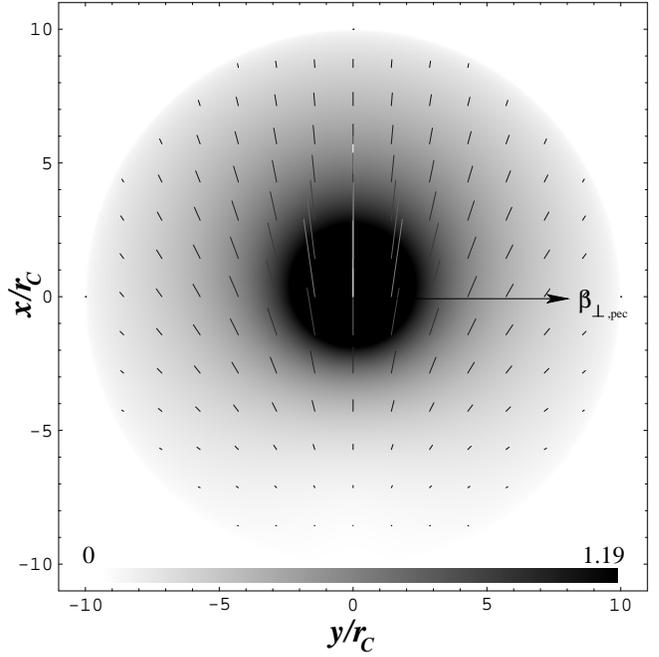}}
\caption{Degree of linear polarisation (grayplot) in units of $G_{\nu}\,\beta^2_{\rm
c}\,\tau_{\rm c}$ for a rotating, oblate cluster ($\iota=1.1$) with cutoff
radius $R=10\,\rC$, an inclination $i=0$ and $\gamma=1.125$ which is moving
with
%---------------------
$\beta_{\perp,\rm b}=10\,\beta_{\rm c}$ in $y$-direction.
%---------------------
The short lines indicate the direction of the planes of polarisation and their
length the degree of polarisation.}
\label{fig:Poli90}
\end{figure}

With Eqns. \eqref{eq:resultQUsup} one finds in this case
%-------------
\beqa\label{eq:Poli0sup} 
P_{i=0} 
=\frac{1}{10}\,G_{\nu,\rm pol}\cdot\beta_{\perp}^2\,
\cdot \tau(x,y,0,\gamma)
\eeqa
%-------------
with 
%-------------
$\beta_{\perp}^2=
\beta_{\perp,\rm b}^2+\beta^2_{\rm c}\,\tilde{r}^2+
2\,\beta_{\perp,\rm b}\,\beta_{\rm c}\,\tilde{x}$.
%-------------
%This result can also be obtained by simply adding the velocity vectors and
%using Eq. \eqref{eq:DPolRJ}. 
The planes of polarisation lie in the direction
%---------------
\beqa\label{eq:PolVeknorm}
\vek{e}_{P}=
\frac{1}{\beta_{\perp}}
\,
\left(
\begin{array}{c}
\beta_{\perp,\rm b}+\beta_{\rm c}\,\tilde{x}\\
\beta_{\rm c}\,\tilde{y}\end{array}
\right)
\Abst{.}
\eeqa
%---------------
If the ICM is not rotating, the planes of polarisation lie along the
$x$-axis. With an additional rotation the planes are tilted by an angle
%---------------
\beqa\label{eq:Tilt}
\cos\varphi&=&\frac{\beta_{\perp,\rm b}+\beta_{\rm c}\,\tilde{x}}{\beta_{\perp}}
\eeqa
%---------------
which can be approximated by
%---------------
\beqa\label{eq:Tiltappr}
\cos\varphi&\approx& 
1-\frac{1}{2}\,\tilde{y}^2\,\delta^2
+\tilde{x}\,\tilde{y}^2\,\delta^3
\eeqa
%---------------
for small ratios $\delta=\beta_{\rm c}/\beta_{\perp,\rm b}$. This can be seen
in Fig. \ref{fig:Poli90}.
%----------------
%----------------

\subsubsection*{Case $i=\pi/2$ and $\alpha=\pi/2$}
%---------------------
The total degree of linear polarisation in this case is the sum of the
contributions due to the rotation and due to the bulk translatory motion
%-------------
\beqa\label{eq:Poli90total} 
P_{i=\pi/2} 
&=&\frac{1}{10}\,G_{\nu,\rm pol}\,\beta^2_{\perp,\rm b}
\cdot\tau(x,y,\pi/2,\gamma)\nonumber\\
& & \qquad\cdot
\left(1+
\frac{\beta^2_{\rm c}}{\beta^2_{\perp,\rm b}}\,
\frac{J(x,y,\pi/2,\gamma)}{\tau(x,y,\pi/2,\gamma)}
\right)
\Abst{.}
\nonumber\\
\eeqa
%-------------
Assuming $\gamma=1$ the relative contribution due to the rk-SZE to the central
degree of polarisation is given as
%-------------
\beqa\label{eq:Poli90cen} 
\frac{P_{i=\pi/2,\rm r}}{P_{i=\pi/2,\rm b}}=
\left[\frac{R}{\arctan R}-1\right]\cdot\delta^2
\Abst{,}
\eeqa
%-------------
where $R$ is the cutoff radius of the cluster. For further discussion see the
following Sect.

\section{Discussion} 
\label{sec:dis}
%----------------
\subsubsection*{Kinetic Sunyaev-Zeldovich effect from cluster rotation}
%---------------------
In Fig. \ref{fig:kinRotClusterInk} the relative change of intensity of the CMB
due to the rk-SZE is shown for an edge on ($i=\pi/2$) view of an oblate
cluster $(\iota=1.1)$ with a cutoff radius $R=10\,\rC$ and $\gamma=1.125$
corresponding to $\beta=3/4$. As mentioned before the distinct dipolar-like
pattern arises from the fact that one part of the gas is moving towards and
the other part is moving away from the observer. If the cluster is viewed face
on ($i=0$) the effect is vanishing, since in that case there are no line of
sight velocity components.

\begin{figure}
\resizebox{\hsize}{!}{\includegraphics{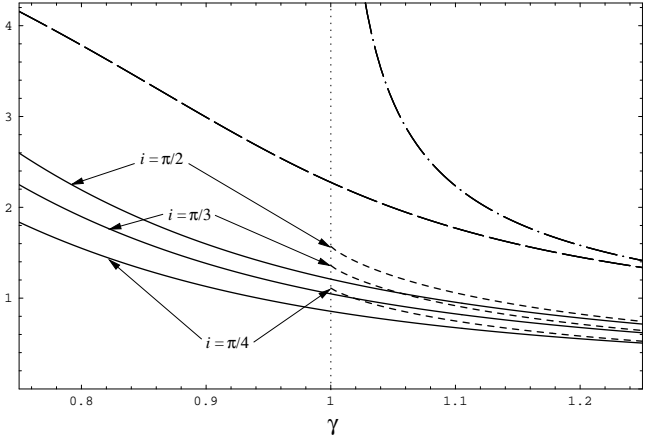}}
\caption{Maximal relative change of intensity $(\Delta I_{\nu}/I_{\nu})_{\rm
rot}$ of a spherical cluster in units of $\tau_{\rm c}\,\beta_{\rm
c}\,G_{\nu}$ for different inclinations $i$: The solid lines correspond to the
case $R=10\,\rC$ and the short-dashed lines to $R=\infty$.  Also the position
$y_{\rm max}$ of the maximal relative change of intensity in units of the core
radius $\rC$ is shown for the case $R=10\,\rC$ (long-dashed) and $R=\infty$
(dashed-dotted). The value of $y_{\rm max}$ does not depend on the inclination
$i$.}
\label{fig:kinRotClusterInkmax}
\end{figure}

Figure \ref{fig:kinRotClusterInkmax} shows the dependence of the maximal
relative change of intensity of the CMB due to the rk-SZE of $\gamma$ for
different inclinations $i$. For increasing $\gamma$ the maximum value
decreases. This is due to the fact that with the increasing steepness of the
electron density profile the main contributions to the intensity change come
from more inner regions of the cluster, which have smaller velocities. In the
case $R\rightarrow\infty$ the relative change of intensity is slightly bigger
than for $R=10\,\rC$. This shows that the contributions of superluminal
velocities far away from the center are negligible. There is no maximum for
$\gamma\le 1$ and $R\rightarrow\infty$. Figure \ref{fig:kinRotClusterInkmax}
also shows the dependence of the position $y_{\rm max}$ of the maximal
relative change of intensity of $\gamma$. For $R=10\,\rC$ it was found
numerically with Eq. \eqref{eq:ResultkinRotClusterInk}. In the case
$R\rightarrow\infty$ we used
%----------
$y_{\rm max}=\eta\AKs{i}\,\rC/2(2\gamma-1)$ 
%----------
from Eq. \eqref{eq:ApprVersch1}. For increasing $\gamma$ the position of
the maximum moves towards the center of the cluster. This is again due to the
increasing steepness of the electron density profile. The position of the
maximum does not depend on the inclination $i$.

%----------------
%----------------

\subsubsection*{Estimates from observational data}
%---------------------
In a recent paper of \cite{Reese2002} the th-SZE at a frequency of $\nu\sim
30\,\text{GHz}$ for a sample of 18 clusters in the redshift range from $z\sim
0.14$ to $z\sim 0.78$ was examined.
%The typical value for the $\beta$-model is about $\bar{\beta}\sim 0.652$
%ranging from $\beta\sim 0.516$ up to $\beta\sim 0.844$.
%---------------------
If we assume that all the clusters in the sample are rotating and they are all
observed edge-on we can estimate the position of the maximal change of
intensity due to the rk-SZE relative to the center of the cluster with the
results for $\beta$ and $\rC$ given in \cite{Reese2002} . The estimates are
given in Table \ref{tab:results}.
%---------------------
In the analysis of \cite{Reese2002} the usual assumption of an
infinite cluster $R\rightarrow \infty$ was made. In this case, with our model
of a rotating cluster there are no distinct extrema for $\gamma\le 1$
(see \eqref{eq:ApprVersch2}). 
%---------------------
The estimates for the case $R=10\,\rC$ correspond to the assumption of an
infinite cluster but a finite region of rotation. A typical value for the
position of the maximal relative change of intensity relative to the center of
the cluster for this sample is $y_{\rm max}\sim 2.6\,\rC$.

With the results of \cite{Reese2002} it is also possible to estimate the
central optical depth $\tau_{\rm c}$. As mentioned in their paper,
relativistic corrections to the th-SZE for their set of clusters are typically
of the order of a few $\%$ in the observed frequency band. Neglecting these
corrections we can make use of Eq. \eqref{eq:FX} and
\eqref{eq:ComptonPar}. Assuming an infinite isothermal $\beta$-model
\citep{Caval76} for the electron density one finds
%---------------
\beqa\label{eq:DT0}
\tau_{\rm c}= 
\frac{2\,\Gamma({\gamma})}{\sqrt{\pi}\,\Gamma({\gamma-\frac{1}{2}})}
\,
\frac{1}{F_{\nu}\,\theta_{\rm e}}\,
\frac{\Delta T_0}{T_{\rm CMB}}\,
\eeqa
%---------------
for the central optical depth, where $\Gamma(x)$ is the Gamma-function and
$\Delta T_0/T_{\rm CMB}$ is the central temperature decrement due to the
th-SZE relative to the CMB temperature $T_{\rm CMB}=2.728\,\text{K}$
\citep{Fixsen96}. The typical central optical depth for the sample of clusters
of \cite{Reese2002} is $\tau_{\rm c}\sim 10^{-3}-10^{-2}$. With the estimates
of the optical depth for each cluster it is possible to calculate the value of
the maximal relative change of intensity due to the rk-SZE. The results are
given in Table \ref{tab:results}. As mentioned before, the results for
$R\rightarrow \infty$ and $R=10\,\rC$ agree well with each other. For the
maximum of $(\Delta I_{\nu}/I_{\nu})_{\rm rot}$ we find a typical value of
$\sim 1.4\,G_\nu\,\beta_{\rm c}\,\tau_{\rm c}$.

\cite{Cooray2002} argue that the velocity at the core radius can be of the
order of a few up to a few ten km/s, that is $\beta_{\rm c}\sim
10^{-5}-10^{-4}$, if there was a recent merger maybe even $\beta_{\rm c}\sim
5\cdot 10^{-4}$ \citep{Roettiger2000,Roettiger1998}. For an observation near the
crossover frequency, $\nu_{\rm c} \sim 217\,\text{GHz}$, where the kinetic and
the rotational kinetic SZE reach their maximum, the relative change of
intensity of the CMB due to the rk-SZE should lead to a temperature change of
the order of
%---------------
$\Delta T_{\rm rot} \sim 10^{-1}-10\,\mu$K,
%---------------
for a recent merger even up to 
%---------------
$\Delta T_{\rm rot} \sim 100\,\mu$K.
%---------------
For the sample of clusters from \cite{Reese2002} the expected temperature
change at the crossover frequency is given in Table \ref{tab:results} assuming
$\beta_{\rm c}\sim 10^{-4}$. It ranges from
%---------------
$\Delta T_{\rm rot} \sim 3.5\,\mu$K 
%---------------
for $\beta_{\rm c}\sim 10^{-4}$ up to
%---------------
$\Delta T_{\rm rot} \sim 146\,\mu$K
%---------------
for a recent merger. Our results lie in the range predicted by
\cite{Cooray2002}.

\subsubsection*{Superposition of all the Sunyaev-Zeldovich contributions}
%---------------------
If the SZE of a rotating cluster is observed, there are in general
contributions of all the different SZEs. The thermal and the kinetic SZE both
peak in the center of the cluster, while the rk-SZE has a maximum and a
minimum displaced symmetrically around the center. We have shown that the
absolute value of relative change of intensity for the superposition of all
the different SZEs has a maximum near the center. For an infinite cluster its
position is approximately at
%----------
$y_{\rm max}=\eta\AKs{i}\,\rC/2(2\gamma-1)$ for small $\eta$,
%----------
where $\eta$ gives the relative contribution of the rk-SZE to
the sum of the thermal and the kinetic SZE and is strongly dependent on the
frequency (see Eq.\eqref{eq:ETA}). 

Observing a cluster at the crossover frequency the parameter $\eta$ becomes
equal to 
%---------------------
$2\,\beta_{\rm c}/\beta_{\parallel,\rm b}$
%---------------------
and is therefore for typical bulk translatory velocities $\beta_{\parallel,\rm b}\sim
10^{-4}-10^{-3}$ of the order $\eta\sim 2\cdot 10^{-2}-2\cdot
10^{-1}$. Assuming an inclination $i=\pi/2$ the displacement of the maximal
relative change of intensity is a few $\%$ of the core radius of the
cluster. For a typical core radius of a few ten arcsec this means a
displacement up to a few arcsec. The displacement is maximal at the crossover
frequency and is negligible in the Rayleigh-Jeans part of the CMB spectrum.
%Future interferometric telescopes by performing multifrequency measurements
%should be able to see this displacement.
Since the main contribution to the peak value of the relative change of
intensity in the considered case comes from the k-SZE, one can in principle
use this effect to measure the rotational velocity component of clusters by
measuring the frequency dependent displacement.
%---------------
\begin{table*}
\centering
\label{tab:results}
\caption{Estimates for $y_{\rm max}$ relative to the center of the cluster,
$(\Delta I_{\nu}/I_{\nu})_{\rm rot}$ and the peak temperature change $\Delta
T_{\rm rot}$ at the crossover frequency for a sample of 18 clusters from
\cite{Reese2002} assuming solid body rotation of the ICM and an inclination
$i=\pi/2$}
%---------------
\begin{tabular}{lccccccc}
\hline\\[-10pt]
& 
\multicolumn{2}{c}{$y_{\rm max}$} & & 
\multicolumn{2}{c}{$(\Delta I_{\nu}/I_{\nu})_{\rm rot}$}
& & $\Delta T_{\rm rot}$
\\[1pt]
& 
\multicolumn{2}{c}{[arcsec]} & & 
\multicolumn{2}{c}{[$G_\nu\,\beta_{\rm c}\cdot 10^{-3}$]}
&  & $[\frac{\beta_{\rm c}}{10^{-4}}\,\mu$K]
\\[2pt]
\cline{2-3} \cline{5-6} \cline{8-8} 
\\[-8pt]
Cluster & $R=10\,\rC$ & $R=\infty$ & & $R=10\,\rC$ & $R=\infty$ & 
& $R=10\,\rC$ 
\\[2pt]
\hline  
\hline\\[-10pt]
\object{MS1137} & 29.2 & 32.4 & & 6.8 & 7.2 & & 7.3\\ 
\object{MS1137} & 29.2 & 32.4 & & 6.8 & 7.2 & &7.3\\ 
\object{MS0451} & 49.6 & 53.7 & & 6.3 & 6.7 & &6.8\\ 
\object{CL0016} & 71.2 & 85.1 & & 8.2 & 9.0 & &8.8\\ 
\object{R1347}  & 26.5 & -- & & 27.3 & -- & &29.1\\ 
\object{A370}   & 156.4 & -- & & 13.8 & -- & &14.8\\ 
\object{MS1358} & 49.8 & -- & & 6.5 & -- & &7.0\\ 
\object{A1995}  & 61.4 & 69.9 & & 5.8 & 6.2 & &6.2\\ 
\object{A611}   & 59.7 & -- & & 8.9 & -- & &9.5\\ 
\object{A697}   & 140.2 & -- & & 10.2 & -- & &10.9\\ 
\object{A1835}  & 37.2 & -- & & 19.9 & -- & &21.2\\ 
\object{A2261}  & 62.5 & -- & & 14.0 & -- & &15.0\\ 
\object{A773}   & 136.3 & -- & & 8.8 & -- & &9.4\\ 
\object{A2163}  & 193.2 & 589.9 & & 8.8 & 11.0 & &9.4\\ 
\object{A520}   & 160.9 & 169.0 & & 3.5 & 3.6 & &3.7\\ 
\object{A1689}  & 76.8 & -- & & 11.4 & -- & &12.2\\ 
\object{A665}   & 202.0 & -- & & 5.1 & -- & &5.4\\ 
\object{A2218}  & 138.9 & 244.8& & 5.7 & 6.7 & &6.1\\ 
\object{A1413}  & 121.7 & -- & & 6.8 & -- & &7.3\\ 
\hline
\end{tabular}
\end{table*}

\subsubsection*{Polarisation due to the kinetic Sunyaev-Zeldovich effect from
cluster rotation}
%---------------------
The degree of linear polarisation of the CMB radiation due to the rk-SZE for
arbitrary inclination of the observer to the rotation axis can be described
with Eq. \eqref{eq:resultPOLrot}.
%---------------------
For the case $i=0$ we showed, that the planes of polarisation for a superposed
kinetic and rotational kinetic SZE are tilted by a angle given by
Eq. \eqref{eq:Tilt}. If we assume that the bulk translatory velocity
$\beta_{\perp,\rm b}$ is 10 to 100 times larger than the rotational velocity
at the core radius $\beta_{\rm c}$, then the planes of polarisation at the
cluster core radius are tilted by an angle of the order of $\sim 0.6^{\circ}$
to $\sim 6^{\circ}$ (see Eq. \eqref{eq:Tiltappr}). This tilt is frequency
independent and can therefore be distinguished from other effects leading to
rotation of the planes of polarisation, like Faraday rotation due to
intracluster magnetic fields.

%---------------------
For a rotating cluster with cutoff radius $R=10\,\rC$ and $\gamma=1$, which
is in addition moving with a bulk translatory velocity $\beta_{\perp,\rm
b}=10\,\beta_{\rm c}$, the contribution of the rk-SZE to the central degree of
polarisation is of the order of $\sim 6\,\%$ for an edge-on observation (see
Eq. \eqref{eq:Poli90cen}). This is equivalent to a temperature change of the
order of
%---------------------
$\Delta T_{\rm r} \sim 10^{-4}\mu$K
%---------------------
and therefore still a challenge for the future. For high precision measurement
of the transversal peculiar velocity component $\beta_{\perp, \rm pec}$ of
clusters the rk-SZE has to be taken into account. 

Here we want to note that the most important contribution to the polarisation
($\sim 0.1 (\tau/0.02)\,\mu$K) arises from the quadrupole component of the CMB
anisotropy \citep{Saz99}. Another important contribution to the polarisation
($\sim 0.01 \,\theta_{\rm e}\,\tau^2$), which can overcome the k-SZE and the
rk-SZE at high frequencies, comes from multiple scattering inside rich
clusters \cite[for a detailed overview see][]{Saz99}. Since the frequency
dependence of all these effects is different they can be separated by
multifrequency measurements.

\section{Conclusions} 
\label{sec:conc}
%----------------
We have derived analytic formulae describing the relative change of intensity
and the degree of linear polarisation of the CMB radiation due to the kinetic
SZE of a rotating cluster of galaxies (Eq. \eqref{eq:ResultkinRotClusterInk}
and \eqref{eq:resultPOLrot}). We also have found analytic formulae for the
superposition of all possible SZE contributions (Eq. \eqref{eq:RELCHANGETOT}
and \eqref{eq:resultQUsup}). We have estimated the possible amplitude of the
relative change of intensity due to the rk-SZE for a sample of 18 clusters of
galaxies in a redshift range from $z\sim 0.14$ to $z\sim 0.78$ (see Table
\ref{tab:results}). Our results show, that the contribution of the rk-SZE to
the peak temperature change for this sample can be expected to range from
%---------------
$\Delta T_{\rm rot} \sim 3.5\,\mu$K 
%---------------
for $\beta_{\rm c}\sim 10^{-4}$ up to
%---------------
$\Delta T_{\rm rot} \sim 146\,\mu$K
%---------------
for a recent merger ($\beta_{\rm c}\sim 5\cdot 10^{-4}$) of rich
clusters. This agrees with the range predicted by \cite{Cooray2002}.

We have also shown that due to the superposition of the thermal, the kinetic
and the rotational kinetic SZE there is a frequency dependent displacement
(Eq. \eqref{eq:Versch}) of the peak value of the relative intensity change,
which in principle can be used to examine the properties of the rotational
velocity component of the ICM. Ground based interferometric telescopes should
be able to detect this effect in the near future. Since the angular momentum
distribution is not easily measurable with other observational techniques,
this aspect of the rk-SZE might provide a new possibility of getting insights
into the internal dynamics of clusters of galaxies.

In the future polarisation measurements of the CMB radiation may become
feasible. In this work we have shown that the polarisation map following from
the k-SZE can alter significantly due to the rk-SZE. Although the contribution
of the rk-SZE to the degree of polarisation relative to the k-SZE is only of
the order of a few $\%$ (quantitatively $\Delta T_{\rm r} \sim 10^{-4}\mu$K),
the directions of planes of polarisation can be tilted by an angle, which is
of the order of a few degrees (Eq. \eqref{eq:Tiltappr}). Since this tilt is
frequency independent it can be easily separated from other effects, for
example Faraday rotation. In the Wien region of the CMB spectrum the degree of
polarisation due to the kinetic and the rotational kinetic SZE increases by a
factor of $10-100$. Therefore measurements of the CMB polarisation should be
performed in this frequency range.

\acknowledgements We acknowledge hospitality at the Universit\"ats-Sternwarte
G\"ottingen, where most of this work was done in the summer of 2001 as part of
the diploma thesis of J.C. We also want to thank S.Y. Sazonov and
T.A. En{\ss}lin for valuable discussions which helped improving the paper.

\appendix
\section{Derivation of the optical depth} 
\label{sec:opticaldepth}
The optical depth is defined as the line of sight integral
%---------------
$\tau=\int\sigT\,\Ne\,dz$
%---------------
over the electron density. Transforming the electron density profile
\eqref{eq:rotClustDichInk} into the frame of the observer and defining the
following abbreviations
%---------------
\bsub\label{eq:Abk1}
\beqa
\xi   &=& bz-\frac{d}{b} \\[1mm]
A   &=& \sqrt{1 + a^2 - \left(\frac{d}{b}\right)^2} \\[1mm]
a^2   &=&\frac{(\AKc{i}^2+\iota^2\AKs{i}^2)x^2 + y^2}{r^2_{\rm c}} \\[1mm]
b^2   &=&\AKs{i}^2+\iota^2\AKc{i}^2 \\[1mm]
d   &=&\frac{\AKc{i}\AKs{i}(\iota^2-1)x}{\rC}
\Abst{,}
\eeqa
\esub
%---------------
where we used $\AKs{i}=\sin{i}$ and $\AKc{i}=\cos{i}$, the optical depth as a
function of $x$ and $y$ is given as
%---------------
\beqa\label{eq:TAU}
\tau(x,y,i,\gamma)
=\frac{\tau_{\rm c}}{\sqrt{\AKs{i}^2+\iota^2\AKc{i}^2}}\cdot I(x,y,i,\gamma)
\Abst{.}
\eeqa
%---------------
Here we defined
%---------------
\bsub\label{eq:Abk2}
\beqa\label{eq:TAUC}
\tau_{\rm c}&=&2\,\sigma_{\rm T}\,n_{\rm e0}\,\rC \\[2mm]
\label{eq:IntdefI}
I(x,y,i,\gamma)&=&\int_0^{\xi_{\rm max}} 
\left(A^2+\xi^2\right)^{-\gamma}\,d\xi\\
\label{eq:XIMAX}
\xi_{\rm max}&=&\sqrt{\frac{R^2}{\rC^2}+1-A^2}
\Abst{.}
\eeqa
\esub
%---------------
The upper limit $\xi_{\rm max}$ results from the boundary of the cluster,
which is parametrized by $R$ assuming an ellipsoidal boundary surface.  The
integral \eqref{eq:IntdefI} can be evaluated analytically and is given as
%---------------
\beqa\label{eq:IntI}
I(x,y,i,\gamma)
=\frac{\xi_{\rm max}}{A^{2\gamma}}
\cdot\,{_2}{F_1}\left(\frac{1}{2},\gamma;\frac{3}{2};-\frac{\xi^2_{\rm max}}{A^2}\right)
\Abst{,}
\eeqa
%---------------
where we made use of the hypergeometric function ${_2}{F_1}(a,b;c;z)$
\citep[see][]{Abram65}.

In the case when $R\rightarrow\infty$ and for $\gamma>1/2$ the integral
$I(x,y,i,\gamma)$ reduces to
%------------
\beqa\label{eq:IntIinfty} 
I_\infty(x,y,i,\gamma) =
\frac{\sqrt{\pi}}{2}
\,\frac{\Gamma\left(\gamma-\frac{1}{2}\right)}{\Gamma(\gamma)} \left( A^2
\right)^{\frac{1}{2}-\gamma}
\Abst{,} 
\eeqa
%------------
where $\Gamma(z)=\int^\infty_0\,t^{z-1}e^{-t}\,dt$ is the Gamma-function.

In the case of $\gamma=1$ the integral $I(x,y,i,\gamma)$ reduces to
%------------
\beqa\label{eq:IntIgamma1} 
I(x,y,i,1) =
\frac{1}{A}\,\arctan\left(\frac{\xi_{\rm max}}{A}\right)
\eeqa
%------------
and has the limiting form $I_\infty(x,y,i,1)=\pi/2 A$ for $R\rightarrow\infty$.

\section{Definition of the integrals $J(x,y,i,\gamma)$ and $H(x,y,i,\gamma)$} 
In the derivation of the degree of linear polarisation due to the rotation of
the cluster medium the integrals
%------------
\bsub\label{eq:DefJH}
\beqa\label{eq:DefJ}
J(x,y,i,\gamma)
&=&\frac{\tau_{\rm c}}{b^3}\cdot I_2(x,y,i,\gamma)
\\
\label{eq:DefH}
H(x,y,i,\gamma)
&=&\frac{\tau_{\rm c}}{b^2}\cdot I_1(x,y,i,\gamma)
\eeqa
\esub
%------------
arise, where we used the abbreviations of appendix \ref{sec:opticaldepth} and
defined the integral $I_n$ as 
%------------
\beqa\label{eq:DefIn}
I_n(x,y,i,\gamma)=
\frac{1}{2}\int_{-\xi_{\rm max}}^{\xi_{\rm max}}\left(\xi+\frac{d}{b}\right)^n
\left(A^2+\xi^2\right)^{-\gamma}\,d\xi
\Abst{.}
\eeqa
%------------
In the limit $R\rightarrow\infty$ the integral $I_n(x,y,i,\gamma)$ diverges
for the combinations $n=2,\,\gamma\leq 3/2$ and $n=1,\,\gamma\leq 1$. For
finite $R$ the solutions
%------------
\bsub\label{eq:JHSol}
\beqa
\label{eq:HSol}
I_1(x,y,i,\gamma)
&=&
\frac{t\,\xi_{\rm max}}{A^{2\gamma}}
\cdot\,{_2}{F_1}\left(\frac{1}{2},\gamma;\frac{3}{2};-\frac{\xi^2_{\rm
max}}{A^2}\right)
\\[1mm]
\label{eq:JSol}
I_2(x,y,i,\gamma)
&=&\frac{1}{3}\,\frac{\xi^3_{\rm max}}{A^{2\gamma}}
\cdot\,{_2}{F_1}\left(\frac{3}{2},\gamma;\frac{5}{2};-\frac{\xi^2_{\rm
max}}{A^2}\right)
\nonumber\\
& &
+\frac{t^2\,\xi_{\rm max}}{A^{2\gamma}}
\cdot\,{_2}{F_1}\left(\frac{1}{2},\gamma;\frac{3}{2};-\frac{\xi^2_{\rm
max}}{A^2}\right)
\eeqa
\esub
%------------
can be found, where we defined $t=d/b$ and made use of the hypergeometric
function ${_2}{F_1}(a,b;c;z)$ \citep[see][]{Abram65}. For $\gamma=1$ and
finite $R$ one finds the solutions
%------------
\bsub\label{eq:JHSol1}
\beqa
\label{eq:HSol1}
I_1(x,y,i,1)
&=&
\frac{t}{A}\cdot\arctan\left(\frac{\xi_{\rm max}}{A}\right)
\\[1mm]
\label{eq:JSol1}
I_2(x,y,i,1)
&=&\xi_{\rm max}-\frac{A^2-t^2}{A}\cdot
\arctan\left(\frac{\xi_{\rm max}}{A}\right)
\Abst{.}
\eeqa
\esub
%------------

\end{document}